# A new method for knowledge representation in expert system's (XMLKR)

Mehdi Bahrami
Payame Noor University (PNU), Tehran, Iran
Mehdi.Bahrami@gmail.com

Dr. Siavosh Kaviani

**Abstract.** Knowledge representation it's an essential section of a Expert Systems, Because in this section we have a framework to establish an expert system then we can modeling and use by this to design an expert system. Many method it's exist for knowledge representation but each method have problems, in this paper we introduce a new method of object oriented by XML language as XMLKR to knowledge representation, and we want to discuss advantage and disadvantage of this method.

## 1. Introduction

Knowledge representation it is an infrastructure of an expert system[5], which introduced many method for this. On the other part, we live in advance and complex information technology world then we need a new method to use of complex knowledge representation.

At first, XML language introduces as extended of HTML language for using, showing and transmitting data of internet webpage[2]. After shortly, this language can be use in other information systems as database system and so.[1]

Nevertheless, the advantage has not been use on this language in expert systems, until now. In this paper at the first section, we survey as background to XML language then discuses total properties of a knowledge representation. Finally, we introduce XMLKR method.

## 2. XML Language

Extensible Markup Language, abbreviated XML, describes a class of data objects called XML documents and partially describes the behavior of computer programs which process them. XML is an application profile or restricted form of SGML, the Standard Generalized Markup Language[7]. By construction, XML documents are conforming SGML documents.

XML documents are made up of storage units called entities, which contain either parsed or unparsed data. Parsed data is made up of characters, some of which form character data, and some of which form markup. Markup encodes a description of the document's storage layout and logical structure. XML provides a mechanism to impose constraints on the storage layout and logical structure.[2]

## 3. Knowledge Representation

Although knowledge representation is one of the central and in some ways most familiar concepts in AI, the most fundamental question about it--What is it?--has rarely been answered directly.[6] What is a knowledge representation? We argue that the notion can best be understood in terms of five distinct roles it plays, each crucial to the task at hand:

A knowledge representation (KR) is most fundamentally a surrogate, a substitute for the thing itself, used to enable an entity to determine consequences by thinking rather than acting, i.e., by reasoning about the world rather than taking action in it.

It is a set of ontological commitments, i.e., an answer to the question: In what terms should I think about the world?

It is a fragmentary theory of intelligent reasoning, expressed in terms of three components: (i) the representation's fundamental conception of intelligent reasoning; (ii) the set of inferences the representation sanctions; and (iii) the set of inferences it recommends.

It is a medium for pragmatically efficient computation, i.e., the computational environment in which thinking is accomplished. One contribution to this pragmatic efficiency is supplied by the guidance a representation provides for organizing information so as to facilitate making the recommended inferences.

It is a medium of human expression, i.e., a language in which we say things about the world.

Understanding the roles and acknowledging their diversity has several useful consequences. First, each role requires something slightly different from a representation; each accordingly leads to an interesting and different set of properties we want a representation to have.

Second, we believe the roles provide a framework useful for characterizing a wide variety of representations. We suggest that the fundamental "mindset" of a representation can be captured by understanding how it views each of the roles, and that doing so reveals essential similarities and differences. [6]

Third, we believe that some previous disagreements about representation are usefully disentangled when all five roles are given appropriate consideration. We demonstrate this by revisiting and dissecting the early arguments concerning frames and logic.

Finally, we believe that viewing representations in this way has consequences for both research and practice. For research, this view provides one direct answer to a question of fundamental significance in the field. It also suggests adopting a broad perspective on what's important about a representation, and it makes the case that one significant part of the representation endeavor--capturing and representing the richness of the natural world--is receiving insufficient attention. We believe this view can also improve practice by reminding practitioners about

the inspirations that are the important sources of power for a variety of representations.[6]

## 4. Semantic Networks

Richard H. Richens of the Cambridge Language Research Unit first invented "Semantic Nets" for computers in 1956 as an "interlingua" for machine translation of natural languages. [9] They were developed by Robert F. Simmons at System Development Corporation, Santa Monica, California in the early 1960s and later featured prominently in the work of M. Ross Quillian in 1966.Semantic networks are knowledge representation schemes involving nodes and links (arcs or arrows) between nodes.[3]

A semantic network is often used as a form of knowledge representation. It is a directed graph consisting of vertices, which represent concepts, and edges, which represent semantic relations between the concepts.

Semantic networks are a common type of machine-readable dictionary.

Important semantic relations:

Meronymy (A is part of B, i.e. B has A as a part of itself)

Holonymy (B is part of A, i.e. A has B as a part of itself)

Hyponymy (or *troponymy*) (A is subordinate of B; A is kind of B)

Hypernymy (A is super ordinate of B)

Synonymy (A denotes the same as B)

Antonymy (A denotes the opposite of B)

An example of a semantic network is WordNet, a lexical database of English.

WordNet involves fairly loose semantic associations, compared to other more formal networks. It is possible to represent logical descriptions using semantic networks such as the Existential Graphs of Charles S. Peirce or the related Conceptual Graphs of John F. Sowa. These have expressive power equal to or exceeding standardfi rst-order predicate logic. Unlike WordNet or other lexical or browsing networks, semantic networks using these can be used for reliable automated logical deduction. Some automated reasoners exploit the graph-theoretic features of the networks during processing.[4]

## 5. Disadvantage of Semantic Networks

Main disadvantage is that the semantics is not clear. It's not clear what the representation means, and so the system doesn't know how to reason or what to infer.[8]

## 6. XMLKR Method

This method include of semantic networks plus XML language that introduce in this paper. As for disadvantage of semantic networks then we have other proposal to knowledge representation, its XMLKR or eXtensible Markup Language for Knowledge Representation.

In this method, at the first phase, we recognize and collect of objects and instance of objects at the world, and draw these items by semantic networks. then at the next phase of drawing for better connection realization and better definition of objects in the world, we use of XML language, by this method we can have a solution to essential problem in semantic networks. Because recognition items and nested relation its essential problem in semantic networks.

In this method, definition and relation between objects and instance of objects use of XML language, but it is only for knowledge representation in expert systems satisfaction.

### 6.1. Object Definition in XMLKR

Figure 1 illustrate how can define Objects, instance of objects and objects properties in the world by XMLKR.

```
<Object 1>
<Object Name> OName </Object Name>
<Attribute 1> Value 1 </Attribute 1>
<Attribute 2> Value 2 </Attribute 2>
.
.
.
<Attribute N> Value N </Attribute N>
<Object 1>
<Object 2>
…
<Object 2>
.
.
.
<Object M>
…
<Object M>
```
Fig1, how to define objects and object attribute

In this way of definition, we can define many arbitrary attribute of an object. In addition, we can define complex structure of objects. For example, figure 2 illustrate of how to define a simple structure in semantic networks (right image) and how to define a similar complex structure in XMLKR(left image).

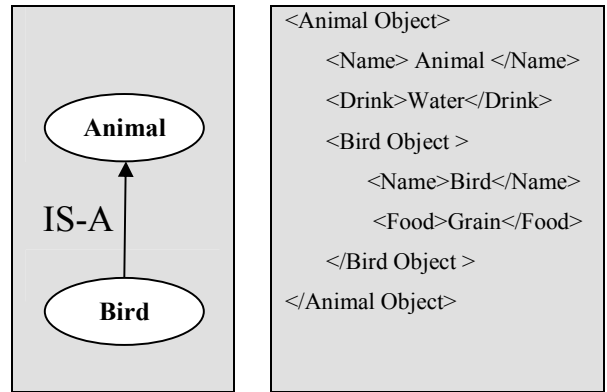

Fig 2, an example of simple semantic networks and similar complex structure in XMLKR

Same as figure 2 attribute of a bird and an animal cannot define simple in semantic networks but we can define simplicity complex structure in XMLKR. In this way, we can define nested structure, for example pursuant of figure 2 we can define grain attribute in bird object or color of grain in grain attribute in bird object.

### 6.2. Objects Relations in XMLKR

In semantics networks, as also explanation typically is two ways for connections:

A. **IS-A**: In this way, we talk about "Has A" and use for inheritance and instance of an object. For description this in XML, we can use of nested object definition as define an attribute of an object. In figure 3, see an object, define by Persia name and it's type of a car object.

```
<Persia Object>
    <Name> Persia </Name>
    <Color>White</Color>
    <ISA Object >
     <Reference>Car
</Reference>
    </ISA Object >
</Persia Object>
```
Fig 3, Example of define an object

B. **AKO**: The various frames are linked together in a hierarchy with a-kind-of (ako) links that allow for inheritance. For example, rabbits and hamsters might be stored in frames that have ako(mammal). In the frame for mammal are all of the standard attribute-values for mammals, such as skin-fur and birth-live. These are inherited by rabbits and hamsters and do not have to be specified in their frames. There can also be defaults for attributes, which might be overwritten by specific species. Legs-4 applies to most mammals but monkeys would have legs-2 specified or specified food at figure 2 for bird by grain.

## 7. Advantage of XMLKR

Attention, to disadvantage of semantic networks, we can introduce advantage of XMLKR as:

- Nomination and definition of objects
- Nomination and definition of complex objects, for define complex object we can use of nested definition
- Capability of define relation for any item
- Complete cover to searching: In semantic networks as discuss later in this paper and discuss in [6] and [7] reference, when define complex object in semantic networks, then we have too time or some when it's go to NP problems, and can't find any polynomial function to solve this problem, but attention to define object and relation in XML language, we have many tools for this as XPath, XSLT and so.[1] for search an item in nested relation object or complex item.

## 8. A challenge in XMLKR and solution

Define loop relation in semantic networks are disadvantage of this type of knowledge representation, because if A object relation to B object and B object relation to C, then C relation to A, then we have a loop in this relation. At the first, we cannot define first object, after this, we cannot define relation between objects, because we must define nested function that do not complete forever. Attention to figure 4:

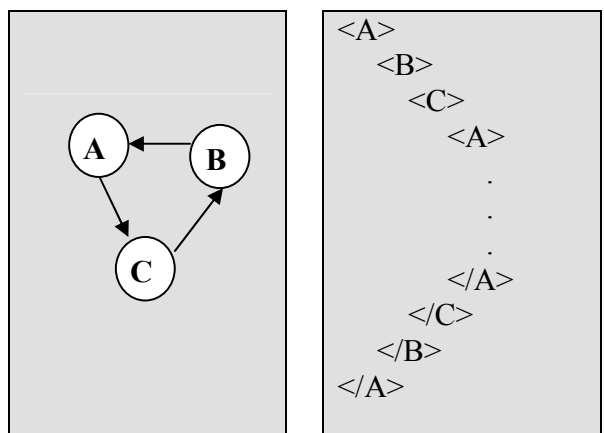

Fig. 4: a loop in Semantic networks, and don't description in XML

Solution to this problem can be solving by below ways:
- Do not use of loop or cycle in relation: This way do not use because loop relation it is need in semantic networks.
- When we want define loop relation, if this connect to first item then do not write again for avoid of loop relation.

For example of figure 4, we can define this relation as figure 5:

```
<A>
    <B>  <C>   </C>    </B>
</A>
<B>
    <C>  <A>   </A>    </C>
</B>
<C>
    <A> <B>  </B>    </A>
</C>
```

Fig. 5: A solution for loop relation in XML

As we see in figure 4, any object distinct defines and if any relations want to connect to the first item do not continue. By this protocol, we can two advantages:
- Object definition
- Search operation in document and find any object

## 9. Conclusion

Semantic networks have many problems for knowledge representation but by new method introduced (XMLKR method) for knowledge representation, we can have solution to all problems of semantic networks and we can define complex object and complex relation.